\documentclass[onecolumn,showpacs,preprintnumbers,amsmath,amssymb]{revtex4}
 \usepackage{graphicx}
\usepackage{latexsym}  
\usepackage{dcolumn}
\usepackage{xcolor}
\usepackage{graphicx, subfigure}

\usepackage{graphicx}% Include figure files
\usepackage{dcolumn}% Align table columns on decimal point
\usepackage{bm}% bold maths 

\usepackage{bm}

\begin{document}

\title{The influence of a strong infrared radiation field on the conductance properties of doped semiconductors}

\author{ I. F. Barna$^{1,2}$, M. A. Pocsai$^{1}$ and S. Varr\'o$^{1,2}$}
\address{1) Wigner Research Centre for Physics of the Hungarian Academy 
of Sciences, \\ Konkoly - Thege Mikl\'os \'ut 29 - 33, 1121 Budapest,  Hungary,\\ 
2) ELI-HU Nonprofit Kft.,  Dugonics T\'er 13, 6720 Szeged, Hungary}
\date{\today}

%%%%%%%%%%%%%%%%%%%%%%%%%%%%%%%%%%%%%%%%%%%%%%%%%%%%%%%%%%%%%%%%%%%%%%%
\begin{abstract}
This work presents an analytic angular differential cross section formula for the electromagnetic radiation field assisted electron scattering by %%  was on
impurities in semiconductors. These impurities are approximated with various model potentials. 
The scattered electrons are described by the well-known Volkov wave function, which  has been used describe strong laser field matter interaction for  more than half a century, %% I would remove this time reference for clarity 
which exactly describes the interaction of the electron with the external oscillating field. 
These calculations show that the electron conductance in a semiconductor could be enhanced by an order of magnitude if an infrared electromagnetic field is present with $ 10^{11} < I < 10^{13}$ W/cm$^2$ intensity.     
\end{abstract}
  
%\draft
\pacs{61.82.Fk,72.20.-i,72.20.Dp,32.80.Wr}
%\keywords{:Semiconductors, Conductivity phenomena in semiconductors and insulators, general theory scattering mechanism, Other multiphoton processes}    
\maketitle

%%%%%%%%%%%%%%%%%%%%%%%%%%%%%%%%%%%%%%%%%%%%%%%%%%%%%%%%%%%%%%%%%%%%%%%                                                  
\section{Introduction}
%%%%%%%%%%%%%%%%%%%%%%%%%%%%%%%%%%%%%%%%%%%%%%%%%
 
The key issue in understanding the electric conduction phenomena in semiconductors is the study  of the corresponding scattering processes of electrons by impurities.
 One possible way to evaluate the rate of scattering transition probabilities is to solve the one-electron  time-dependent Schr\"odinger equation up to the first Born approximation. Numerous models exist to approximate the electron-impurity interaction via a central potential of $ U({\bf{r}})$ and the standard description can be found in textbooks  \cite{karlheinz,shang,davies,solyom}. This study extends this description to the case where an electromagnetic (EM) field is simultaneously present in addition to impurity induced scattering. 

The main motivation for this work comes from the field of laser-matter interactions. In this topic the non-linear response of atoms, molecules and plasmas can be both investigated theoretically and experimentally \cite{muller}. These result in many phenomena including high harmonic generations or plasma-based laser-electron acceleration. 
 The original theory of potential scattering in external EM fields was developed about half a century ago and can be found in various papers of \cite{bunk,bunk1,faisal,kroll,gont,bergou,berg_varro,faisal2}.
Numerous studies  on laser assisted electron collisions on atoms are also available \cite{kam}.   
Kanya and Yamanouchi generalized the Kroll-Watson formula \cite{kanya} for a single-cycle infrared pulse and applied it to time-dependent electron diffraction.
There are only two studies where heavy particles e.g.,  protons,  are scattered by nuclei in strong electromagnetic fields \cite{barna1,barna2}.  
Similar theoretical studies of solid states or semiconductors in such strong electromagnetic fields are rare and recently,  it has became possible to investigate the band-gap dynamics \cite{schul} and the strong-field resonant dynamics \cite{wis} of semiconductors in the attosecond (as) time scale.

This paper contains a self-contained overview of electron conduction calculations in a doped semiconductor; the  theory of laser assisted 
potential scattering and the numerical calculation of Lindhard dielectric function -- all of which are essential tools for the presented theoretical description. 

Finally, numerical calculations were performed for a model potential in infrared electromagnetic fields with intensities $  10^{11} < I < 10^{13}$ W/cm$^2$. 
The photon energy of such fields are below 1~eV which is comparable to some semiconductor band gaps. 
It was shown by Kibis \cite{kibis} that the backscattering of conduction electrons are suppressed by strong high-frequency electromagnetic field and this effect does not depend on the absolute form of the scattering potential.  
Later, Morina {\it{et al.}} \cite{morina} calculated the transport properties of a two-dimensional electron gas 
which interacts with light. This can be considered as the precursor of this study. 

This investigation shows that electrical conductivity of doped semiconductors can be changed by more than a magnitude by the presence of a strong infrared radiation field. This may open the way to a new forms of electronic gating. These kind of coherent infrared radiations will be soon available e.g., at the ELI-ALPS Research institute in Szeged, Hungary \cite{eli}. 

%%%%%%%%%%%%%%%%%%%%%%%%%%%%%%%%%%%%%%%%%%%%%
\section{Theory} 
%%%%%%%%%%%%%%%%%%%%%%%%%%%%%%%%%%%%%%%%%%%%%%%%
%%%%%%%%%%%%%%%%%%%%%%%%%%%%%%%%%%%%%%%%%%%
\subsection{Electron scattering on impurities in semiconductors}
%%%%%%%%%%%%%%%%%%%%%%%%%%%%%%%%%%%%%%%%%
A short overview of the derivation, using first quantum mechanical  and statistical physical principles, of the electron conductivity without the laser field is now presented.
A much detailed derivation can be found in many theoretical solid state physics textbooks e.g., \cite{karlheinz,shang,davies,solyom}. 

Free electrons are considered in three dimensions with the initial and final states defined as plane waves with the following form
\begin{equation}
\varphi_i({\bf{r}}) =  \frac{A}{(2\pi \hbar)^{3/2}} exp \left(    \frac{i}{\hbar}{\bf{p_i}}\cdot{\bf{r}}\right), \hspace*{1cm} \varphi_f({\bf{r}}) =  \frac{A}{(2\pi \hbar)^{3/2}} exp \left(    \frac{i}{\hbar}{\bf{p_f}}\cdot{\bf{r}}\right). 
\label{wave}  	 
\end{equation}
The considered perturbation is simply due to the extra potential energy of the impurity $U({\bf{r}})$ and therefore the transition rates can be evaluated as 
\begin{equation}
U_{fi} = \int \varphi_f({\bf{r}})^*U({\bf{r}}) \varphi_i({\bf{r}})d{\bf{r}} = \frac{1}{A}\int U({\bf{r}})e^{-i {{\bf{q\cdot r}}}}d^2 {\bf{r}}. 
\end{equation}
This is the two-dimensional Fourier transform of the scattering potential and ${\bf{q}} = {\bf{p_i}} - {\bf{p_f}}$ is the momentum transfer of the scattering electron, where $ {\bf{p_i}} $ and $ {\bf{p_f}}$  stand for initial a final electron momenta.  
The differential Born cross section of the corresponding potential is given by 
 $ \frac{d \sigma_B}{d \Omega} = \left(\frac{m}{2\pi \hbar^2} \right)^2 |U({\bf{q}})|^2$.  %% this should be an equation for clarity!

The well-known total scattering cross section $\sigma_T$ for the elastic process can be calculated from the differential scattering cross section via an angular integration, where the 
back scattered electrons gives significant contributions therefore  a $[1-cos(\theta)]$  factor appears
\begin{equation}
\sigma_T = 2\pi \int_0^{\pi} \left( \frac{d \sigma_B(\theta)}{d\Omega}    \right)   [1-cos(\theta)]sin(\theta) d\theta. 
\label{szigma}
\end{equation}
The relaxation time or the $\tau$ single-particle life time against impurity scattering is defined in terms of the total scattering cross section  by multiplying with the number of impurities $n_{imp}$ 
\begin{equation}
1/\tau = \sigma_T n_{imp}. 
\label{relax}
\end{equation}
Finally the electron mobility and the conductivity are  defined by
\begin{equation}
\mu = e\tau/m_e ,  \hspace*{1cm} G = e \mu n_e
\end{equation}
where $e,m_e,n_e$ are the elementary charge, effective mass and  the number of the scattered electrons, respectively.   
Further technical details including references and an overview over various additional methods  e.g., the derivation of  Eq. (3) from Boltzmann equations, are given in the review of  Chattopadhyay \cite{chatt}.
This model is only valid for "dilute" semiconductors  where the concentration of the doping atoms is below a given threshold and thus, the effects of multiple scattering can be neglected. For silicon this value lies around $10^{15}$ cm$^{-3}$, the degeneracy level is at  $10^{18}$ cm$^{-3}$.  %%% It isn't clear what these two numbers relate too, is Silicon typicall 10-15cm-3 and the dopant level of .1%?? Needs to be rephrased
%%%%%%%%%%%%%%%%%%%%%%%%%%%%%%%%%%%%%%%%%%%%%%
\subsection{Electromagnetic field assisted potential scattering}
%%%%%%%%%%%%%%%%%%%%%%%%%%%%%%%%%%%%%%%%%%%%%%%%%%

The following section contains a detailed summarization of the non-relativistic quantum mechanical description of this system.  
The coherent infrared field is treated semi-classically via the minimal coupling. %minimal coupling what? approach, model approximation.. it needs to be a something 
The IR beam is taken to be linearly polarized and the dipole approximation is used. 
The non-relativistic description in dipole approximation is only valid of the dimensionless intensity parameter (or the normalized vector potential)  $a_0 = 8.55 \cdot 10^{-10} \sqrt{ I (\frac{W}{cm^2})} \lambda (\mu m)$  of the external field is smaller than unity. A laser wavelength of 3 $\mu m$ correlates to a critical intensity of $ I = 1.52 \cdot 10^{17} $ W/cm$^2$, however much smaller laser intensities will be assumed and a moderate electron kinetic energy below one eV will be considered. 

To avoid the ionization of the lattice atoms, silicon will be considered to have a band gap of 1.12 eV and  a 3 $\mu$m infrared (IR) electromagnetic field has  a photon energy of 0.41 eV. 
Therefore, the ionization of  the highest energy bound valence electron would require a three photon absorption process in the perturbative regime. 
The probability of absorbing $N$ photons depends on the laser intensity, $I$, as $I^N$.  
The characteristic field strength in an atom are rather high and  correspond to a laser intensity of   $ 3.5 \cdot 10^{16} $  W/cm$^2$.  In this work, it is considered that   
IR field intensities $I = 10^{12}$  W/cm$^2$, which is much lower than the atomic units. 

The following Schr\"odinger equation has to be solved in order to describe  the non-relativistic scattering process of an electron on an impurity by an external EM field 
\begin{equation}
\left[ \frac{1}{2m} \left(  {\bf{\hat{p}}}- \frac{e}{c} {\bf{A}} \right)^2 + U({\bf{r}}) \right]\Psi  = i \hbar  \frac{\partial \Psi}{\partial t}, 
\label{sch}
\end{equation}
where ${\bf{\hat{p}}}= -i\hbar \partial/\partial {\bf{r}}$ is the momentum operator of the electron and $U({\bf{r}})$ represents  the scattering potential of the impurity atom,  ${\bf{A}}(t)= A_0 {\bf{e}} cos(\omega t)$ is the vector potential of the radiation field with unit polarization vector of ${\bf{e}}$. 
Figure 1 shows the geometry of the scattering event. 
Without the scattering potential $ U({\bf{r}})$ the particular solution of (\ref{sch}) can be immediately written down as 
non-relativistic Volkov states  $\varphi_p({\bf{r}},t)$ which exactly incorporate the interaction with the EM field, 

%%%%%%%%%%%%%%%%%%%%%%%%%%%%%%%%%%%%%%%%
\begin{figure}
%[!h]
\begin{center}
%* \vspace*{1.0cm} 
%\hspace*{0.5cm}
\scalebox{0.4}{
\rotatebox{0}{\includegraphics{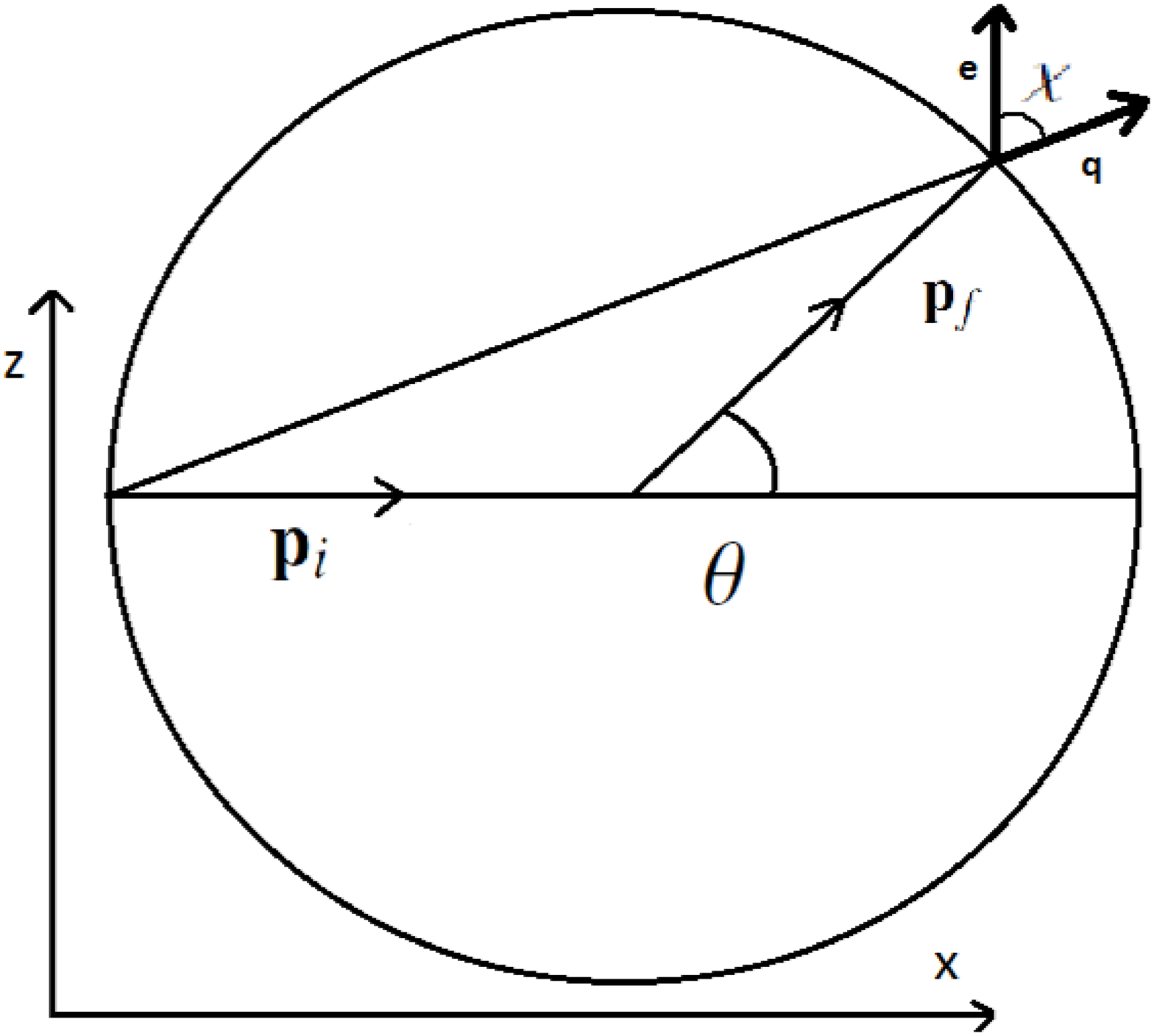}}}
\end{center}
\caption{The geometry of the scattering process. The impurity atom is located in the center of the circle,  
${\bf{p}}_i$ and  ${\bf{p}}_f$ are the initial and final scattered electron momenta, $\theta$ is the 
electron scattering angle, the EM pulse propagates parallel to the x axis and is linearly polarized in the x-z plane. 
The $\chi$ angle is needed for calculating the EM-electron momentum transfer.}  	
\label{egyes}       % Give a unique label 
\end{figure}
%%%%%%%%%%%%%%%%%%%%%%%%%%%%%%%%%%%%%%%%
%where 
The Volkov states are modulated de Broglie waves; parametrized by momenta ${\bf{p}}$ and form an orthonormal and complete set
\begin{equation}
\begin{split}
\varphi_p({\bf{r}},t) =&  \frac{1}{(2\pi \hbar)^{3/2}} exp\left[    \frac{i}{\hbar}{\bf{p}}\cdot{\bf{r}}  - \int_{t_0}^t dt'  \frac{1}{2m}
\left(  {\bf{p}} - \frac{e}{c} {\bf{A}}(t')  \right)^2 \right], 
\end{split}	 	 	
\label{volk}
\end{equation}

\begin{equation}
\int d^3 r \varphi_p^*({\bf{r}},t) \varphi_{p'}({\bf{r}},t)  = \delta_3 ({\bf{p}}- {\bf{p'}}) , \hspace*{1cm}
\int d^3 p \varphi_p({\bf{r}},t) \varphi_{p}^*({\bf{r'}},t)  = \delta_3 ({\bf{r}}- {\bf{r'}}).
\label{norm}
\end{equation}

To solve the original problem of Eq. (\ref{sch}) the exact wave function are written as a superposition of an incoming Volkov state 
and a correction term, which vanishes at the beginning of the interaction (in the remote past $t_0 \rightarrow -\infty$). 
The correction term can also be expressed in terms of the Volkov states as these form a complete set (see Eq. (\ref{norm})), 

\begin{equation}
\Psi({\bf{r}},t) = \varphi_{p_i}({\bf{r}},t) + \int d^3p a_p(t) \varphi_p ({\bf{r}},t), 
\hspace*{1cm} a_p(t_0) =0. 
\label{nagy}
\end{equation}

It is clear that the unknown expansion coefficients $a_p(t) $ describe the non-trivial transition symbolized as ${\bf{p}}_i \rightarrow {\bf{p}}$, from a Volkov state of momentum ${\bf{p}}_i$ 
to another Volkov state with momentum {\bf{p}}. The projection of $\Psi$ into some Volkov state $\varphi_p(t)$ results in  

\begin{equation}
\int d^3 r \varphi_p^*({\bf{r}},t) \Psi({\bf{r}},t)  = \delta_3 ({\bf{p}}- {\bf{p}}_i) + a_p(t). 
\label{norm2}
\end{equation}

The insertion of  $\Psi$ of Eq. (\ref{nagy}) into the complete Schr\"odinger equation (\ref{sch}) results in the following integro-differential equation for the coefficients $a_p(t)$, 

\begin{eqnarray}
 i\hbar \dot{a}_{p'} (t) = 
\int d^3 r  \varphi_{p'}^*({\bf{r}},t') U({\bf{r}})  \varphi_{p_i}({\bf{r}},t') +  \nonumber \\ 
\int d^3 p  a_p(t) \int d^3 r \varphi_{p'}^*({\bf{r}},t') U({\bf{r}})  \varphi_{p}({\bf{r}},t'),  
\end{eqnarray}

where the scalar product was taken with $\varphi_{p'}(t)$ on both sides of the resulting equation 
and the orthogonality property of the Volkov sates was taken after all (see the first Eq. of (\ref{norm})). 
The initial condition $a_p(t_0)=0$, 
already shown in (\ref{norm}), %% ?? Not sure what is meant here
means that the formal solution of (\ref{sch}) can be written as 

\begin{eqnarray}
 a_{p'} (t) = 
-\frac{i}{\hbar}  \int_{t_0}^t  dt'  \int d^3 r  \varphi_{p'}^*({\bf{r}},t') U({\bf{r}})  \varphi_{p_i}({\bf{r}},t')  \nonumber \\
 -\frac{i}{\hbar}  \int_{t_0}^t  dt'
\int d^3 p  a_p(t') \int d^3 r \varphi_{p'}^* ({\bf{r}},t') U({\bf{r}})  \varphi_{p}({\bf{r}},t').  
\label{inte}
\end{eqnarray}

In the spirit of the iteration procedure used in scattering theory the $(k+1)-$th iterate of $a_p(t)$ is 
expresses by the k-th iterate on the right hand side in  (\ref{inte}) as 

\begin{eqnarray}
 a_{p} ^{(k+1)}(t) = 
-\frac{i}{\hbar}  \int_{t_0}^t  dt'  \int d^3 r  \varphi_{p'}^*({\bf{r}},t') U({\bf{r}})  \varphi_{p_i}({\bf{r}},t')  \nonumber \\ 
 -\frac{i}{\hbar}  \int_{t_0}^t  dt'
\int d^3 p  a_p^{(k)}(t') \int d^3 r \varphi_{p'}^*({\bf{r}},t') U({\bf{r}})  \varphi_{p}({\bf{r}},t').  
\label{inte2}
\end{eqnarray}

In the  first Born approximation, where the transition amplitude is linear in the scattering potential $U({\bf{r}})$ , the transition amplitude has  the form

\begin{equation}
T_{fi} = \lim_{t \rightarrow \infty}\lim_{t_0 \rightarrow -\infty} a^{(1)}_{p_f}(t) = 
-\frac{i}{\hbar}  \int_{-\infty}^{\infty}  dt'  \int d^3 r  \varphi_{p_f}^*({\bf{r}},t') U({\bf{r}})  \varphi_{p_i}({\bf{r}},t').
\label{melem}
\end{equation}

The $A^2$ term drops out from the transition matrix element  (\ref{melem}), because it represents a uniform time-dependent phase.  
By taking the explicit form of the Volkov states (\ref{volk}) with the vector potential  $
A(t) = {\bf{e}} A_0 cos(\omega t) $ means that  $T_{fi}$ 
becomes 
\begin{equation}
 T_{fi} = \sum_{n=-\infty}^{\infty} T_{fi}^{(n)} , \hspace*{1cm} 
 T_{fi}^{(n)}  = -2\pi i \delta \left(   \frac{p_f^2-p_i^2}{2m} + n \hbar \omega  \right)
J_n(z) \frac{ U({\bf{q}})}{(2\pi \hbar)^3},     
\label{scat_form}
\end{equation}
Before time integration, the exponential expression can be expanded into a Fourier series with the help of the Jacobi-Anger formula \cite{abr}.
This results in the next formula including the Bessel functions of the first kind 
\begin{equation}
e^{izsin(\omega t)} = \sum_{n=-\infty}^{\infty} J_n(z)e^{in \omega t}. 
\end{equation}

The $U({\bf{q}})$ is the Fourier transform of the scattering potential with the momentum transfer of 
 ${\bf{q}}  \equiv {\bf{p}}_i  - {\bf{p}}_f  $ where  ${\bf{p}}_i$ is the initial and 
${\bf{p}}_f$ is the final electron momenta its absolute value is $    q = \sqrt{p_i^2 + p_f^2 - 2p_ip_f cos(\theta_{p_i,p_f})}$. 
In the case where electrons have  0.1 - 1 eV energy in the $n=0$ channel (which means elastic scattering), the following approximation is valid $q \approx 2 p_i  |sin(\theta/2)| $.

In general, the Dirac delta function describes photon absorptions $(n<0)$ and emissions $(n>0)$.  
$J_n(z)$ is the Bessel function with the argument depending on the parameters of the laser field, the intensity and the frequency     
$z   \equiv   \frac{2a_0 q sin(\theta/2) cos(\chi)}{\hbar \omega /c} $
where  $a_0, q, \chi $ are the dimensionless intensity parameter, the momentum transfer of the electron and the angle between the momentum 
transfer and the polarization direction of the EM field, respectively. 

  The general differential cross section formula for the laser assisted collision with simultaneous nth-order photon absorption and stimulated emission processes are  
\begin{equation}
\frac{d \sigma^{(n)}}{d \Omega} = \frac{p_f}{p_i}J_n^2(z) \frac{d \sigma_B}{d \Omega}. 
\label{cross}
\end{equation} 

The $ \frac{d \sigma_B}{d \Omega} = \left(\frac{m}{2\pi \hbar^2} \right)^2 |U({\bf{q}})|^2$ 
is the usual Born cross section for the scattering on the potential U(r) alone, without the external EM field. 
The expression Eq. (\ref{cross}) was calculated by several authors using different methods \cite{bunk,bunk1,faisal,kroll,gont,bergou,berg_varro,faisal2} and  if the Born cross section is exactly known, Eq. (\ref{cross}) can be substituted in Eq. (3)  and the single-particle lifetime can be easily calculated. 
 
%%%%%%%%%%%%%%%%%%%%%%%%%%%%%%%%%%%%%%%%%
\subsection{Scattering model potentials in semiconductors}
%%%%%%%%%%%%%%%%%%%%%%%%%%%%%%%%%%%%%%%%%%%%%%%%%%%

Different kinds of analytic model potentials are available to model the electron scattering on impurities in a semiconductor and four are presented within this paper. The simplest model is the "box potential" which is well-known from quantum mechanics textbooks.  %%ref?
It is capable to mimic the two-dimensional impurity scattering provide by a cylindrical-barrier of radius $a$.  It  can be used to describe a neutral impurity such as an Aluminum atom that has diffused from a barrier into a GaAs well \cite{davies}.  

Mathematically,  $U(r) = U_0$ if $r \le a$ and $U(r)= 0$ if $r > a,$ 
where $U_0$ is the depth of the square well potential in eV and $a$ is the radius in nm. 
The two dimensional Fourier transformation of the potential gives the first-order Bessel function of the form of 
$ U(q) = \pi a^2U_0 \frac{J_1(2qa)}{qa} $.  
The infinite range Coulomb potential has an infinite total scattering cross section in the first-order Born approximation. 
However,  considering a maximal limiting impact parameter  due to electron screening in semiconductors the isotropic elastic cross section can be
 evaluated (\ref{szigma}) and done by \cite{conwell}. %% I don't understand this bit, it doesn't quite make sense
This is the Brooks-Herring(BH) model \cite{brooks,herring} and is applicable to describe electron scattering on an ionized impurity atom  
\begin{equation}
U(r) = \frac{\varepsilon e^{-r/\lambda_D}}{4\pi \epsilon_0 \epsilon_r r}, 
\end{equation}
where $\lambda_D = \sqrt{ \frac{\epsilon_0 \epsilon_r k_B T}{q^2 n_0} }$ 
is the Debye screening length, $\epsilon_0$ and $\epsilon_r $  are the vacuum permitivity and the dielectric constant of the present media. 
To avoid confusion $\varepsilon$ is used for the charge of the ionized impurity atom instead of $q$ which is fixed for the momentum transfer of the electrons.  
The Debye screening is just the solution of the linearized Poisson-Boltzmann theory and is the simplest way to handle the problem. More general theory that describe screening would be Landau's approach known as Fermi liquid theory where there is a quantitative  account of electron-electron interaction \cite{aschr}.  

The Fourier transform of this potential is as follows \cite{shang},
\begin{equation}
U(q) =  \frac{\varepsilon^2\lambda_D^2}{\epsilon_0\epsilon_r(1+ q^2 \lambda_D^2)}. 
\label{BH}
\end{equation}
More realistic screening lengths can be calculated with the Friedel sum and the phase shift analysis of the potential \cite{chatt}. There are numerous models available from the original BH interaction  
which include various additional effects e.g., dielectric of Thomas-Fermi screening, electron-electron interaction \cite{chatt}. 
The original BH can also be calculated from a realistic electron concentration of the ionized impurity containing the Fermi-Dirac 
integral \cite{dingle}.   

There are two additional potentials which are widely used to model impurities in semiconductors. The first has been developed to investigate the electron charged dislocation scattering in an impure electron gas. The derivation of the formula 
can be found in \cite{karlheinz}. The radial potential has the form of 
$
U(r)= \frac{\epsilon}{2\pi \epsilon_r c}K_0\left( \frac{r}{\lambda_D}\right)
$
where $K_0$ is the zeroth-order modified Hankel function, $\epsilon,\epsilon_r, 1/c, \lambda_D$ 
are the elementary charge, dielectric constant, linear charge density and the Debye screening 
length.  This dislocation is a two dimensional interaction and has a cylindrical symmetry.  
The Fourier transform of the potential is   
$
U(q) = \frac{ \epsilon \lambda^2}{ \epsilon_r c (1+ q^2 \lambda^2)}. 
$
Jena \cite{jena0} used this interaction to evaluate the quantum and classical scattering times due to charged dislocation in an impure electron gas. 
Half a century earlier, P\"od\"or \cite{pod} calculated an analytic formula for relaxation time and investigated the 
electron mobility in plastically deformed germanium. These are remarkable similar to the three dimensional BH potential. 

The last most advanced model is dipole scattering in polarization  induced two-dimensional electron gas. This considers the electrical field of a dipole above a plane \cite{jena1}.    
In the following,  the BH model will be analyzed in details.  

%%%%%%%%%%%%%%%%%%%%%%%%%%%%%%%%%%%%%%%%%%%%%
\subsection{Generalized field assisted potential scattering in a media}
%%%%%%%%%%%%%%%%%%%%%%%%%%%%%%%%%%%%%%%%%

The previouly outlined laser assisted potential scattering description with the listed potentials is not sufficient for a realistic model to evaluate electron conduction in a solid at finite temperature. 
Thereforer two additional improvements are considered. 

Firstly, the scattering electrons now move in a media (doped semiconductor) instead of a vacuum, therefore the effect of the media, 
the dielectric response functions, has to be taken into account. 
%% up to a level of complexity. ???
Eq. (\ref{cross}) is modified and generalized  and the numerically evaluated Lindhard dielectric function \cite{lind} is included in the scattering potential. 
It can be shown, using quantum Vlasov theory,  in the first Born approximation using the Wigner representation of the 
density matrix of the electron \cite{kull} that the total interaction potential in the frequency domain is equivalent to the Fourier 
transform of the interaction potential in vacuum multiplied by the Lindhard dielectric function. Therefore 
the final differential cross section formula is
\begin{equation}
\frac{d\sigma^{(n)}}{d\Omega} = \frac{p_f}{p_1}\left(  \frac{m}{2\pi\hbar^2} \right)^2 J_n^2(z) | U({\bf{q}}, \epsilon_r[k,\omega]) |^2. 
\end{equation}
The dielectric function now depends on the angular frequency of the external applied field, the coherent IR field,  and the wave vector of the scattering electron. 
The correct form of the interaction for the BH model is : 
\begin{equation}
U(q,k,\omega) =  \frac{\varepsilon^2\lambda_D^2}{\epsilon_0\epsilon_r(k,\omega) (1+ q^2 \lambda_D^2)},   
\label{corBH}
\end{equation}
 where  $ \varepsilon$ in the numerator is the charge of the impurity.  
The next technical step in the model is to calculate the Lindhard dielectric function.  
For a fermion gas with electronic density $n$ at a finite temperature,
$T$, the form can be expressed in terms of real and imaginary part \cite{arista} 
\begin{equation}
\varepsilon_r(k,\omega) = \varepsilon_{r_{R}}(k,\omega) +  i\varepsilon_{r_I}(k,\omega). 
\label{Lind}
\end{equation}
At finite temperatures, %% or a finite temperature
 the dielectric function contains singular integrals of the Fermi function which can be eliminated with 
various mathematical transformations. According to \cite{anc},  the following expressions have to be evaluated: 
\begin{equation}
\varepsilon_{r_{R}}(k,\omega) = 1 + \frac{1}{4\pi k_F \kappa^3} [g_t(\lambda_+ = u + \kappa) - g_t(\lambda_- = u -\kappa)],
\label{real_eps}
\end{equation}
and 
\begin{equation}
\varepsilon_{r_I}(k,\omega) = \frac{t}{8 k_F \kappa^3}\ln \left[ \frac{1+exp(\alpha(t) -\lambda_-^2)/t}{1+exp(\alpha(t) -\lambda_+^2)/t}  \right].
\label{im_eps}
\end{equation} 
 Where the Fermi wave vector is $k_F = [3\pi^2n]^{1/3}$, the reduced temperature is $t= T/T_F$, the Fermi energy is 
$E_F = k_F^2/2 = k_BT_F$  (where the electron mass and $\hbar$ were set to unity).  
The reduced variables $u$ and $\kappa$ introduced by Lindhard are defined as 
\begin{equation}
 u = \frac{\omega}{v_F k}, \hspace*{1cm} \kappa = \frac{k}{2k_F},
\end{equation} 
where $\omega$ is the angular frequency of the IR field and $k$ is the wave vector or the scattered electron in the model. 
First, the reduced chemical potential $\alpha(t) = \mu/E_f$ has to be evaluated at a finite temperature %% or finite temperatures
from the integral of 
\begin{equation}
\int_0^{+\infty} x^2 \frac{1}{1+\exp( \frac{x^2-\alpha(t)}{t} )}dx = \frac{1}{3}.
\end{equation} 
After determining the chemical potential, the function $g_t(\lambda)$ can be calculated via an integral where the usual singularity is successfully eliminated by a appropriate mathematical transformation 
 \begin{equation}
g_t(\lambda) = \lambda^2 \int_0^{\infty} \left[-2A \frac{X \exp(AX^2 -B)}{1+\exp(AX^2-B)^2}   \right] \times 
\left[  -X + \frac{1-X^2}{2}\ln \left| \frac{X+1}{X-1}\right| \right] dX.
\end{equation}
where $A = \lambda/t$ and  $B = \alpha(t)/t$.
Exhaustive technical details can be found in the original paper \cite{anc}. 

At this point, the static $\epsilon(q, \omega \rightarrow 0)$ and the long wavelength limit  $\epsilon(q\rightarrow 0,\omega)$ of the Lindhard function can be reduced to analytic formulas \cite{lind,aschr}.  For the static limit $\epsilon(q,0) =  1 + \frac{\kappa^2}{q^2} $, the 3D screening wave number,$ \kappa $, (3D inverse screening length) is defined as 
$\kappa = \sqrt{\frac{4\pi e^2 }{ \varepsilon} \frac{\partial n}{\partial \mu} }$ where $n, \mu,\varepsilon$ are the particle density $N/L^3$, chemical 
potential and charge, respectively. However, in the long wavelength limit in 3D, $\epsilon(0,\omega) = 1 - \frac{\omega^2_{plasma}}{\omega^2}$ 
where the angular frequency of the plasma reads $ \omega^2_{plasma} = \frac{4\pi e^2 N}{\varepsilon L^3 m}$. 

At a finite temperature %% or finite temperatures
in a realistic semiconductor,  the scattering electrons are not monochromatic and thus an averaging over the distribution has to be evaluated  
\begin{equation}
\langle G \rangle = \frac{e^2 n_e}{m^{*}n_{imp}  \langle \sigma_T \rangle }.
\label{tau_av}
\end{equation} 
This means that there is an additional numerical integration of the total cross section multiplied by the Fermi-Dirac distribution function 
$f(E)$ (for non-degenerate electrons) times the density of states $g(E)$ according to Shang \cite{shang},     
\begin{equation}
\langle \sigma_T \rangle = \frac{\int_0^{\infty}  \sigma_T(E)  f(E) g(E) dE  }{ \int_0^{\infty} f(E) g(E) dE }
\label{aver}
\end{equation} 
with $E(k) = \frac{\hbar^2k^2}{2m}$ being the energy of the electrons.  
In this representation, the integration can be traced back to an integral over $k$. 
The numerical value of the density of state function is well-known for one, two or three dimensional solids. 
A two dimensional system $g(E)_{2D} = \frac{m_e}{2\pi \hbar^2} $  is independent of the electron energy. In three dimensions, the current BH potential case,   $g(E)_{3D} = \frac{m_e^{3/2}}{\sqrt{2}\pi^2\hbar^2} \sqrt E $.  The mentioned charged dislocation potential is a two dimensional model.

If the number of donors are enhanced, the Fermi level will rise towards the conduction band. At some stage 
the approximations will no longer hold because 
%%more than only the tail %% this sounds wrong to me
a larger proportion %% this sounds more scientific, is it true?
of the Fermi Dirac function overlaps with the band edge. 
The approximations break down when the Fermi level is closer than  $3  k_{B}T$  to one of the band edges. This is approximately 75 meV at room temperature.
In this case, the semiconductor become degenerate and the Boltzmann distribution function has to be applied, instead of the Fermi-Dirac fucntion. 
To the best of our knowledge, these two completions were never added to the general laser assisted potential scattering to model electron scattering in realistic solid states.  

In practical calculations the upper limit of the integral can be cut at the Fermi energy which is about 1~eV at room temperature for semiconductors.  
Numerical values obtained from (\ref{aver}) with or without external electromagnetic field can be directly compared in the future to experimental data. 

%%%%%%%%%%%%%%%%%%%%%%%%%%%%%%%%%%%%%%%%%
\section{Results} 
%%%%%%%%%%%%%%%%%%%%%%%%%%%%%%%%%%%%

Doped  silicon has been considered as a semiconductors with a Fermi energy of 1~eV and  coherent IR electromagnetic sources ($\lambda$ = 1 -- 5 $\mu$m) are  the external field. According to the ELI-ALPS' white book \cite{white},  a $\lambda = 3.2$  $\mu$m wavelength mid-IR laser will operate,%% This Laser is here and working!! I would reference the Cornier paper!  This makes the paper seem old and out of date
which will be  similar to \cite{26}. 
An intensity range of $ 10^{12 } <I < 10^{13} $ W/cm$^2$ is used to avoid laser damage of the silicon sample.   
Stuart {\it{et al.}} \cite{stuart} and  Tien {\it{et al.}} \cite{mur1}  published experimental results for radiation damage of silica for  $\lambda = 1$  $\mu$m wavelength laser pulses with different pulse length and found that the threshold lies at $10^{13}$ W/cm$^2$ for 100 fs pulse duration.  
Unfortunately, no experimental measurement values for $\lambda = 3 \mu$m could be found. 
However, there is an empirical power law dependence for damage threshold for silica glass at various wavelength  $I_{th}(\lambda) = 1.55 \cdot \lambda^{0.43} \cdot I_{th} $ , where the wavelength and the intensities should be given in $\mu$m and W/cm$^2$ units \cite{kochner}. This means that the threshold  at 3 $ \mu m$  should be  $ 2.5 \cdot 10^{13}$ W/cm$^2$. 
 
According to \cite{chatt}, 6 electron mobility versus electron concentration measurement were presented and compared to various BH models on a logarithmic-linear scale below 200 K giving discrepancy of a factor of $ 2 - 5 $ which validates the original electron conduction model. These results were the encouragement to develop the above mentioned model. %% why isn't this in the introduction or somewhere earlier. It feels out of place here.

The lowest level of the kinetic energy of the electrons considered was  thermal noise $E = k_BT$ which is 0.025 eV at room temperature. 
The parameters of the BH potential are the following the screening range $\lambda_d = 30$ nm,  the dielectric constant $\epsilon_r = 35$ and the 
charge  $\varepsilon =1$. 

In the following, calculations for two distinct models are presented.
The first system  simply considers a semiconductor media with a dielectric constant.  The literature value is 35 %%units?
for semiconductor Silicon.  %This is the simplest way to take into account the effect of silica. 
The second  more realistic model fully includes the frequency dependent Lindhard function Eq.  (\ref{Lind}).  
This model even includes the energy dependence of the scattering electrons.  The two field independent cross sections are 
$\sigma_{T,\epsilon = 35} = 1.9$ nm$^2$ and $\sigma_{T,\epsilon(\omega,k)} = 39.6$ nm$^2$, respectively. The $\epsilon(k)$ dependence and averaging over the electron energy makes the ratio %% what is this ratio? needs to be defined - not clear
 less than 35 -- the numerical value is 20.8.  
 
Figure 2. shows the averaged cross sections as the function of the external field intensity for $\lambda= 3$ $  \mu$m wavelength. 
The more realistic model%%, which includes the Lindhard function --- You've litterally just said this, removed for clarity
 gives larger cross sections which means that there is a smaller electron conductance.  
By considering the field independent cross section $\sigma_{T,\epsilon = 35} = 1.9 $~nm$^2$ as a standard value gives the suppression of the electron 
conductance by the external field by a factor of 15.   
 
Fig 3. presents the ratio of the two models as the function of the field intensity with values between 21 and 26. At larger field intensities the gradient of the ratio is reduced.   

Figure 4. shows the averaged cross sections as the function of the external field wavelength for the intensity of  $I = 10^{12}$   W/cm$^2$. 
The cross sections obtained from the model incorporating the Lindhard function are still larger than the simpler model. The cross sections of both models decay at large intensities.   
Note, that for 1 $\mu$m wavelength, the ratio of the original cross section (1.9 nm$^2$) goes up to 310 nm$^2$ %% how has this ratio got units and others not? check!
a gain factor of 155.  
Calculations below  1 $\mu$m wavelength were not performed  because such fields may excite valence electrons into the conductance band and that would lie out of the scope of this elastic scattering model.   
Figure 5 shows the wavelength dependence of the ratio of these two models.  
Larger wavelength means smaller cross sections or larger conductance. These results are in fully agreement with the general theory of \cite{kibis}.
The ratio between the two models still lie at a factor of 23 to 25. %%units????   
  
Multiplying with the remaining constants of $\frac{e^2 n_e}{m_e n_{imp}}$ where $e$ is the charge of the electron,   
$m_e$ is the effective mass of an electron in a semiconductor is about $0.5 \times 9.0\cdot 10^{-31}$ kg and the number of the impurities per cm$^3$ 
lies between $10^{9} - 10^{16}$ therefore the obtained final conductance values would lie between $10^{-6} - 10^{5}$ Si/cm \cite{milton}.  %% I have no idea what this setence is trying to say. It needs to be written
This is a very broad range of conductance, hence not reporting exact numerical conductance values. However, these ratio of the conductances with or 
without strong external IR fields can vary by more than a magnitude.    

A doped semiconductor has a complex nature and the physical value of the resistivity change in an external IR field can of course only be 
investigated in a real physical experiment but these calculation shows that it would be an interesting project. 

These models only include elastic scattering processes, without any photon absorption or emission. The inclusion of  one photon 
absorption only requires changing the zeroth order Bessel function to the first order term, otherwise the process and the way of calculation are the same. Note, that the corresponding cross sections or the probabilities of a first order process is at least one order of magnitude lower than elastic one.   

These presented calculations cannot include additional effects coming from the complex nature of a real solid state like, valence dielectric screening, 
band-structure details, electron-electron scattering, non-linear screening, multiple electron scattering and impurity dressing -- all of which are mentioned in the review of \cite{chatt}.

The two-fold numerical integrations of (\ref{aver}) for various laser parameters were evaluated with Wolfram Mathematica 
[Copyright $1988 - 2012$ Wolfram Research, Inc.] where the global adaptive integration built-in method was used with recursion number of 300. 
Numerical precision was set up to ten digits.  %% this seems really out of place here? At the begining?

%%%%%%%%%%%%%%%%%%%%%%%%%%%%%%%%%%%%%%%%%%%%%%%%%%%%
\begin{figure}
%%* \vspace*{1.0cm} 
%%\hspace*{0.5cm}
\scalebox{0.5}{
\rotatebox{0}{\includegraphics{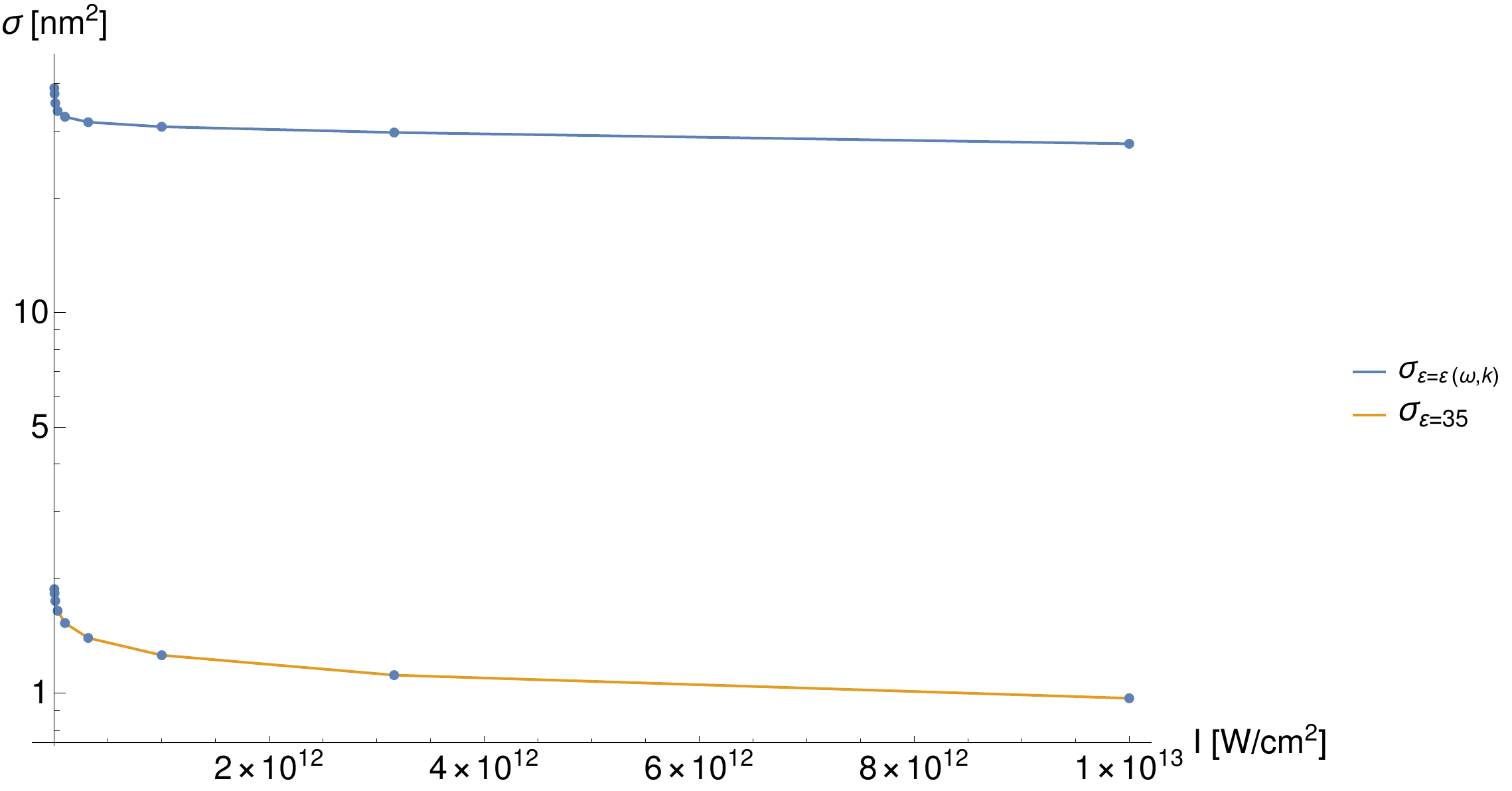}}}  
\caption{The averaged cross section $\langle \sigma_T \rangle $ Eq. (\ref{aver}) as the function of the field intensity for a 
$\lambda = 3 \mu$m  laser wavelength  at room temperature. 
The  upper curve is for the frequency dependent Lindhard dielectric function. The lower curve is for the  $\epsilon_r = 35$ dielectric 
constant case.  }  
\label{F2}       % Give a unique label of 
\end{figure}
%%%%%%%%%%%%%%%%%%%%%%%%%%%%%%%%%%%%%%%%%%%%%%%%%%%%%% 
%%%%%%%%%%%%%%%%%%%%%%%%%%%%%%%%
\begin{figure}
%%* \vspace*{1.0cm} 
%%\hspace*{0.5cm}
\scalebox{0.4}{
\rotatebox{0}{\includegraphics{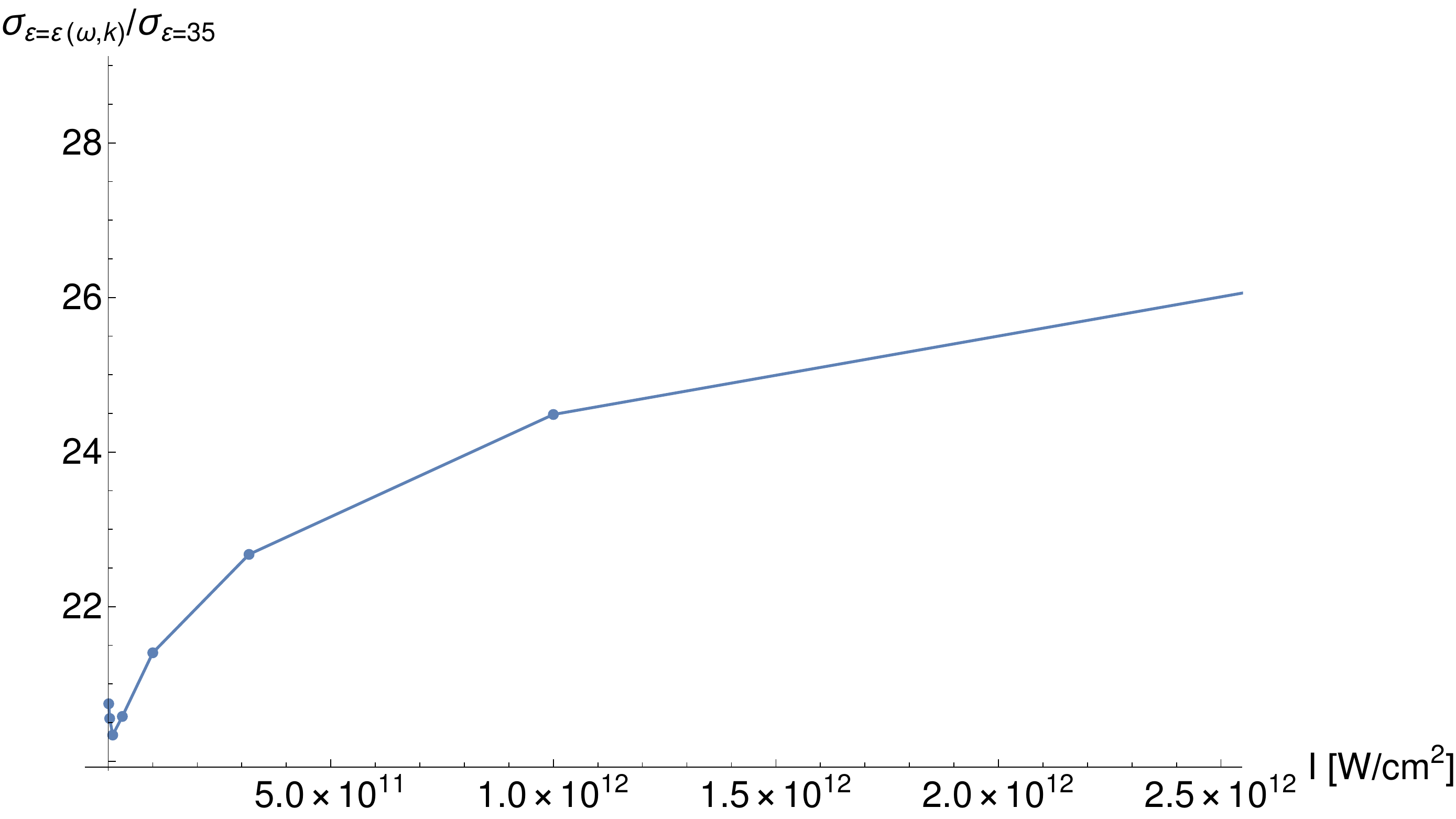}}}  
\caption{The intensity dependence of the ratio of the two models at $\lambda = 3 $ $\mu$m.}    
\label{F3}       % Give a unique label of 
\end{figure}
%%%%%%%%%%%%%%%%%%%%%%%%%%%%%%%%%%%

%%%%%%%%%%%%%%%%%%%%%%%%%%%%%%%%
\begin{figure}
%%* \vspace*{1.0cm} 
%%\hspace*{0.5cm}
\scalebox{0.5}{
\rotatebox{0}{\includegraphics{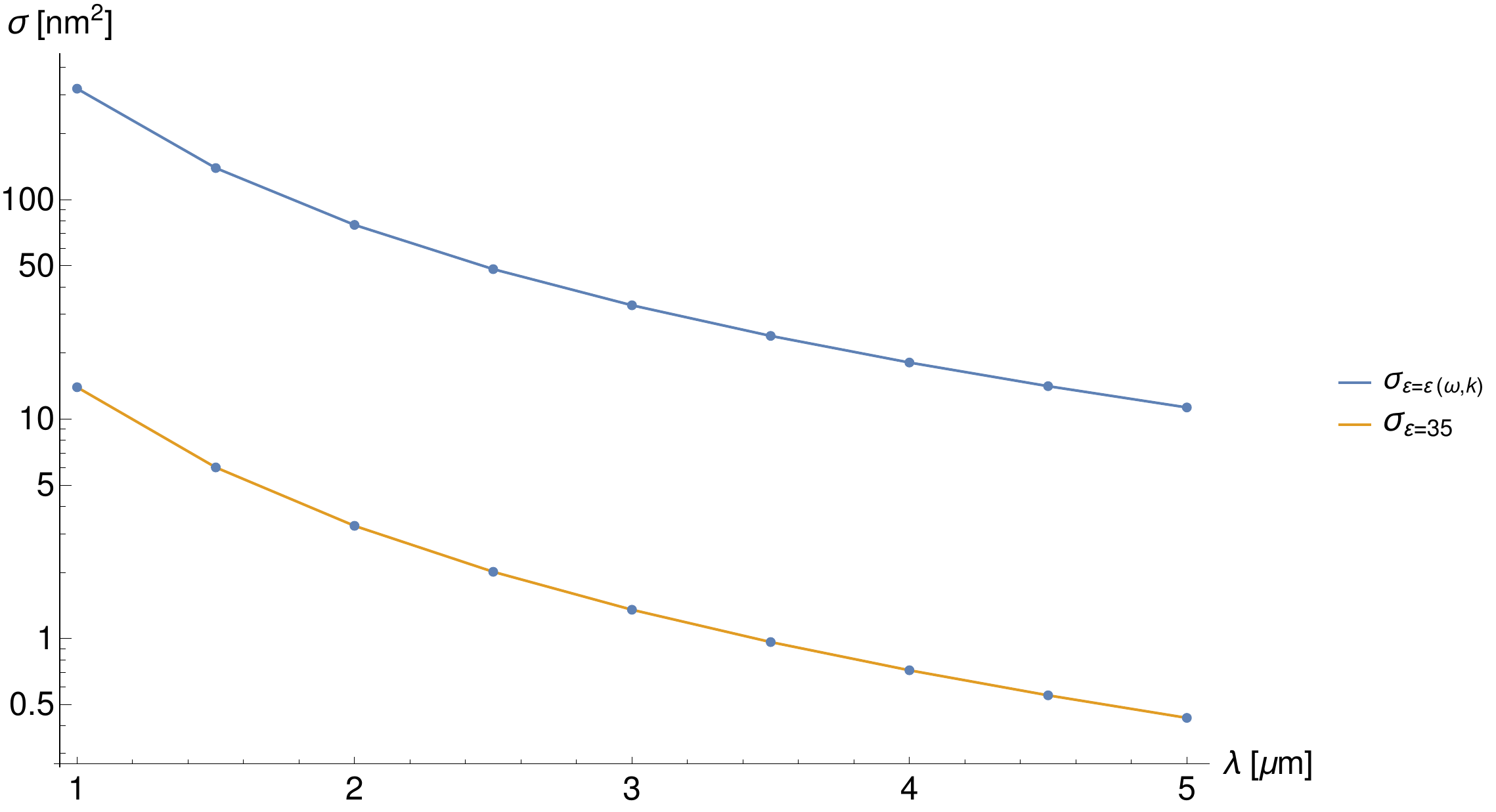}}}  
\caption{The averaged cross section $\langle \sigma_T \rangle $ Eq. (\ref{aver}) as the function of the wavelengths for  
$ I = 10 ^{12}$   W/cm$^2$ field intensity at room temperature.
The  upper curve is for the frequency dependent Lindhard dielectric function. The lower curve is for the  $\epsilon_r = 35$ dielectric 
constant case. }  
\label{F4}       % Give a unique label of 
\end{figure}
%%%%%%%%%%%%%%%%%%%%%%%%%%%%%%%%%%%

%%%%%%%%%%%%%%%%%%%%%%%%%%%%%%%%%%%%%%%%%%%%%%%%%%%%
\begin{figure}
%%* \vspace*{1.0cm} 
%%\hspace*{0.5cm}
\scalebox{0.4}{
\rotatebox{0}{\includegraphics{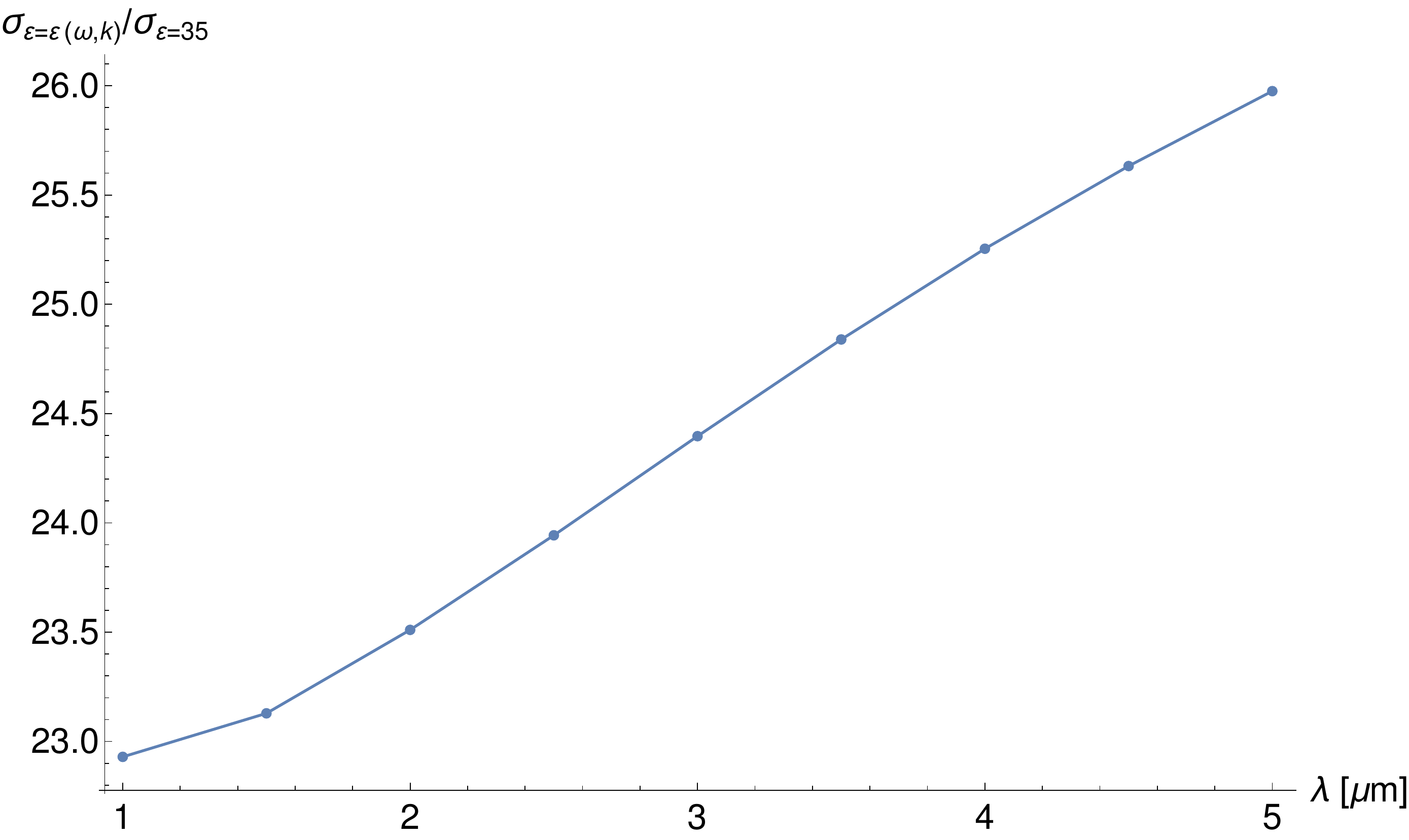}}}  
\caption{The wavelength dependence of the ratio of the two models at $I = 10 ^{12} $ W/cm$^2$ intensity. } 
\label{F5}       % Give a unique label of = 
\end{figure}
%%%%%%%%%%%%%%%%%%%%%%%%%%%%%%%%%%%%
\section{Summary} 
A formalism based on an interrelated model to calculate electron conductance in a doped semiconductor in strong external 
IR fields with intensities between $10^{11} - 10^{13}$ W/cm$^2$ with a wavelength range of $1 --  5$ $\mu$m.  
The mathematical description of multi-photon processes was coupled to the well-established potential scattering model based on the first-Born approximation.
An application of this  general formalism in the present paper has been  the modification of the scattering elastic cross sections in the elastic channel.   

In solid sate physics, the elastic scattering of electrons on impurities modeled by the BH potential can model the electric conductance up to a factor 
of $ 2 $ to $5 $, a reliable background \cite{chatt}. Electron scattering has been treated in a perturbative manner whilst  the influence of the external strong radiative IR field (which can cause photon absorption) has been treated non-perturbatively.  The solid state scattering model has been improved in two areas.  A  frequency dependent  Lindhard function which mimics the response of the solid to a quick varying external field has replaced the inclusion of dielectric constant, which models the semiconductor. The second improvement is that the final electron energy distribution above the Fermi function at room temperature is averaged. These two improvements give at least a factor of 15 suppression in the final conductance. 
It has been demonstrated that due to the joint interaction of the conduction electrons with the impurity scattering potential and the laser field that  there could be  a considerable change in the conduction as was expected at the beginning of this studies.  
These theoretical results have inspired our experimental colleagues in ELI-ALPS to perform measurements on silicon samples at 3 $\mu$m wavelength. Work is in progress to build up a setup to measure the prognosticated change in the conductivity. If the change of the conductivity lies in the same order of time scale as the duration of the mid-IR laser pulse (tenth of femtoseconds) than only pump-probe type measurements can be applied to measure the change of  the optical properties of the sample \cite{hugen}.  
If the experiments verify these theoretical predictions than it may be possible to start speculating about possible physical application of the phenomena, like a quick gating, a quick moldulator or even a mid-IR light intensity sensor. 
 
\section{Acknowledgments} 
The Authors would like to thank Dr. Ugo Ancarani for the useful discussions about numerical problems that arose from calculating Lindhard functions.   
The ELI-ALPS project (GINOP-2.3.6-15-2015-00001) is supported by the European Union and co-financed by the European
Regional Development Fund.
      
%%%%%%%%%%%%%%%%%%%%%%%%%%%%%%%%%%%%%%%%%%%%%%%%%%%%%%%%%%%%%%%%%%%%%%                                    
                                                                  
%%%%%%%%%%%%%%%%%%%%%%%%%%%%%%%%%%%%%%%%%%%%%%%%%%%%%%%%%%%%%%%%%%%%%%%        
\end{document}